\documentclass[%
preprint,secnumarabic,prfluids,
nobibnotes,
 amsmath,amssymb,
 aps,
]{revtex4-2}

\usepackage{graphicx}
\usepackage{dcolumn}
\usepackage{bm}
\usepackage{placeins} 
\usepackage{xcolor} 

\newcommand{\modif}[1]{{#1}}

\begin{document}

\preprint{APS/123-QED}

\title{Experimental control of Tollmien-Schlichting waves using the Wiener-Hopf formalism}

\author{Diego B. S. Audiffred}
 \email{diegodbsa@ita.br}
\author{André V. G. Cavalieri}%

 \author{Pedro P. C. Brito}%
\affiliation{%
 Instituto Tecnológico de Aeronáutica, Brazil\\
}%


\author{Eduardo Martini}
\affiliation{Institut Pprime - CNRS - Université de Poitiers - ISAE-ENSMA, France \\
}%

\begin{abstract}

Reactive flow control has been shown as a promising tool to improve, among other aspects, the aerodynamic characteristics of an aircraft. This paper focuses on the use of reactive flow control to attenuate Tollmien-Schlichting (TS) waves over a wing profile. TS waves are an instability mechanism that is one of the first stages of boundary layer transition to turbulence. The Wiener-Hopf technique was used in this work for the experimental boundary layer control. The approach improves previous wave-cancellation techniques that, by constructing control kernels in the frequency domain, lead to control kernels with a non-causal part, i.e., actuation would need future sensor information to be constructed. In practical applications, it is unfeasible to access this type of information. Ignoring the non-causal part of the kernel leads to suboptimal solutions that might significantly degrade the performance of the controller. The Wiener-Hopf formalism allows us to take into account causality constraints in the formulation of the control problem, leading to an optimal realistic solution and a control kernel that is causal by construction. Moreover, it is possible to construct the control strategy based only on the power and cross-spectra obtained experimentally, in a data-driven approach. The present work shows \modif{how to apply experimentally the Wiener-Hopf resolvent-based formalism using signals from a wind tunnel experiment, demonstrating} that the Tollmien-Schlichting waves can be effectively attenuated via a Wiener-Hopf based controller, which yielded better results than a typical wave-cancellation approach. 

\cleardoublepage
\end{abstract}

\maketitle


\section{\label{sec:intro} Introduction}

 Turbulent flows are associated with higher friction drag levels compared to the laminar state, which implies more fuel consumption, and, in turn, higher aircraft operational cost. It is worth mentioning that reducing $CO_2$ emissions is an imperative task for the aircraft industry; for example, the Advisory Council for Aviation Research and Innovation in Europe (ACARE) has established the goal of reducing the $CO_2$ emission by 75\% by 2050, compared to the values observed in 2000 \cite{acare}.

Flow control approaches have been present in the aviation field for many decades. However, only recently reactive flow control has been seen as a promising tool to the development of safer and more efficient aircraft. Nowadays, it is considered for a wide range of applications in the aircraft industry, such as aeroacoustics \cite{Maia2021}, aerodynamic load reduction \cite{Williams2018}, to suppress the buffet instability \cite{gao2017}, to eliminate the adverse aerodynamic effects of icing on wings \cite{POURYOUSSEFI2016}, for flight control \cite{Warsop2018}, but mostly to reduce the skin friction drag \cite{Abbas2017,Varshney2020, Brito2021}, which may be achieved by delaying the boundary layer transition to turbulence.

For a flat plate, or an unswept wing, in a free stream with low turbulence intensity ($Tu < 0.1 \%$, where $Tu$ is the turbulence level), two-dimensional instability waves are generated in the boundary layer by external disturbances. Such instability waves are known as Tollmien-Schlichting (TS) waves; they are the primary instability mechanism of the transition process and can be described by the linear stability theory (LST). TS waves grow exponentially as they travel downstream, and once they reach an amplitude of about 1\% of the free-stream velocity, a secondary instability mechanism leads to three-dimensional disturbances that forms $\Lambda \textrm{- vortices}$. These vortices will form turbulent spots that will coalesce into a fully turbulent flow \cite{Schlichtin2017,Kachanov1994}.

While Tollmien-Schlichting waves occupy an extensive part of the transition region, the subsequent stages are more complex and occur more rapidly. These characteristics of the transition process, and by the fact that TS waves grow by a linear mechanism, make the TS-growth region a natural choice for the use of control strategies aiming to delay the transition. For that reason, the majority of the studies dedicated to transition delay of boundary layer, focus on the attenuation of TS waves, in situations where these are the primary instability mechanism. 

The first experiments related to the attenuation of Tollmien-Schlichting waves were reported by \citet{Schilz1965} and \citet{Wehrmann1965}. Both had investigated the attenuation of TS waves over a flat plate. where the former had shown that the imposition of an acoustic field may either lead to an earlier transition or delay it. He also used flexural waves to the excitation and suppression of TS waves. The latter used a vibrating ribbon to generate TS waves at a single frequency that were suppressed by an active wall. More than a decade later, \cite{Milling1981} performed experimental control of TS waves, where a vibrating wire was used to generate a TS wave type disturbance over a flat plate placed in a water channel, with the same principle being used for the control mechanism, where a second wave, with a $180^{\circ}$ phase shift with respect to the original disturbance, was used to reduce the amplitude of the TS wave. Nowadays, reactive control approaches still use wave-cancellation principles for the attenuation of Tollmien-Schlichting waves \cite{sasaki2018, Brito2021}. Other reactive control techniques that are used for this purpose include: Linear Quadratic Gaussian (LQG) \cite{semeraro2013,Tol2019}, Proportional–Integral–Derivative (PID) \cite{Li2021}, Filtered-x Least-Mean-Squares Algorithm (FxLMS) \cite{STURZEBECHER2003, fabbiane2015,Kotsonis2015} and Model Predictive Control (MPC) \cite{Ghiglieri2012}. The last three techniques are essentially capable of dealing with parameter uncertainty of the flow control system, however, when compared with the LQG, for instance, the robustness of such methods costs a drop in performance for the design condition.

The linear quadratic regulator (LQR) is a well known control strategy used for linear systems, where an optimal solution for the controller is obtained when a quadratic cost functional is minimized with respect to the kernel, which leads to a Riccati equation. Typically, the use of a reduced order model (ROM) is required to lower the degrees of freedom of the system, allowing the solution of the Riccati equation, since the order of fluid dynamic systems is usually too high. Furthermore, LQR relies on the full knowledge of the system state. Thus, the estimation of the full state from a restricted number of noisy measurements is often required. This might be performed with a Kalman filter, which when combined with the LQR, leads to optimal estimation and control. For such case, the controller is referred to as linear quadratic Gaussian, or LQG. An additional Riccati equation is necessary to solve in order to obtain an estimation kernel.

A wave-cancellation approach is a more straightforward technique and prevents the need of using ROMs. A further advantage is the possibility to use directly transfer functions educed from experiments, in a data-driven approach \cite{Brito2021}. Within this context, for a feed-forward scheme, a controller can be obtained from the direct inversion of transfer functions in the frequency domain. In the literature, this technique is also referred to as inverse feed-forward control (IFFC) \cite{Brito2021, sasaki2018}. For such case, the controller is obtained in the frequency domain, with the actuation signal calculated as follows:

    \begin{equation}
       \mathbf{u}(\omega) = \mathbf{\Gamma}(\omega)\mathbf{y}(\omega) 
       \label{eq:K_IFFC}
    \end{equation}

\noindent where $\mathbf{u}$ is the actuation signal, $\mathbf{y}$ is the reference sensor signal and $\Gamma$ is the control kernel. In the time domain, this will result in the convolution given by:

    \begin{equation}
       \mathbf{u}(t) = \int ^{\infty} _{-\infty} \mathbf{\Gamma}(\tau)\mathbf{y}(t-\tau)d\tau
       \label{eq:kt_IFFC}
    \end{equation}

\noindent which might lead to a solution where the actuation $\mathbf{u}$ depends on negative values of $\tau$. In a concrete application that is not feasible, the actuation signal would require future sensor readings information. In these cases, the control is denominated as non-causal. 

A causal control \modif{can be based on} past sensor measurements \modif{only} ($\tau>0$ in Eq. (\ref{eq:kt_IFFC}))\modif{. IFFC acomplishes that truncating the non-causal kernel, i.e., setting $\Gamma(\tau<0)=0$.} However, this approach might substantially decrease the performance of the controller, including for the specific case of TS wave attenuation \cite{Brito2021, Martini2022}.

\modif{An alternative to IFFC is to construct an ROM from the measurement data, and then use the LQG framework to obtain a causal control law \cite{morra2020, semeraro2013, Tol2019}.} We propose the use of the Wiener-Hopf formalism\modif{, which, in this work, is constructed based on acquired data only, } for the experimental control of Tollmien-Schlichting waves. Under the Wiener-Hopf approach, an optimal causal solution is obtained for a controller solved in the frequency domain, which is equivalent to the solution that would be obtained by using LQG \modif{\cite{Martini2022, Martinelli2009, Martinelli2009paper}}.


\modif{The Wiener-Hopf framework, although well established in the control literature, has rarely being applied for flow control. Only a few studies have been conducted in this regard. \citet{Martinelli2009} and \citet{Martinelli2009paper} were the first reported studies that considered the Wiener-Hopf technique for flow control applications, where it was used for control of wall turbulence in a channel flow, aiming at drag reduction. A rate of dissipation norm, obtained with Fourier expansion in the spatial homogeneous directions of a parallel or quasi-parallel flow, has been used to build the objective functional. They showed that such approach allows the design of the optimal feedback controller in frequency domain, and with only one single step. In the LQG approach, the design of the optimal controller is obtained with the Kalman filter providing the necessary state estimate, so the LQG method requires two steps to solve the control problem, as states the Separation Theorem \cite{AndersonMoore1990}. It was shown that this frequency-domain approach is computationally more efficient than LQG, especially when the number of sensors and actuators are relatively small if compared to the dimension of the state matrix.} 

\modif{No other flow control study has been developed in this regard until more recently, when \citet{Martini2022} constructed the Wiener-Hopf regulator from linearized equations of motion and a model of the forcing using a resolvent-based framework. The potential of the method was illustrated for an estimation and control problem of a flow over a backwards-facing step. The obtained results showed the potential of the method to improve wave-canceling strategies by minimizing the effect of kernel truncation. These studies were compared against the Kalman filter, where \citet{Martinelli2009} obtained correlation coefficients between the actual state and the estimate states for each method. The correlation coefficients obtained were considered ``strikingly similar". \citet{Martini2022} showed that the results obtained with the Wiener-Hopf approach are equivalent to those obtained with LQG. The Wiener-Hopf approach yielded the same estimation and control gains obtained with the Kalman filter and the LQG. For further details about the relationship about these two estimation and control approaches, the reader is referred to the works of \citet{Martinelli2009} and \citet{Martini2022}.}

\modif{These previous studies dealt with control problems in numerical simulations,} so, to the best of our knowledge, this is the first time the Wiener-Hopf technique is used in a flow control application facing the challenges intrinsic to experiments. In particular, we wish to determine the ability of the Wiener-Hopf method to attenuate TS waves, using spectra and transfer functions obtained directly from experiments. This, on one hand, ensures that one deals with a close approximation of the system at hand, minimizing modeling errors; on the other hand, this exposes the control design to the unavoidable experimental noises and errors. The TS-wave control experiment previously developed by our group \cite{Brito2021}, where control was carried out using IFFC, is  conducted using Wiener-Hopf controllers in this work, in order to show potential advantages of the latter method. 

\section{\label{sec:methods} Methodology}

\subsection{Experimental setup}

Experimental boundary layer control over a NACA 0008 airfoil was performed in an open-circuit wind tunnel with a test section of  1.28 m x 1.0 m and turbulence level inferior to $0.1 \%$. The airfoil, with 0.8 m chord, was manufactured by SAAB and the KTH Royal Institute of Technology. A pexiglass plate at one of the airfoil surfaces was used in order to accommodate pressure sensors and a plasma actuator. The airfoil was positioned at zero angle of attack with respect to an undisturbed freestream flow of 10 m/s, which yielded a chord based Reynolds number of $5.33\cdot 10^5$.

\begin{figure}[!ht]
\includegraphics[width=0.95\textwidth]{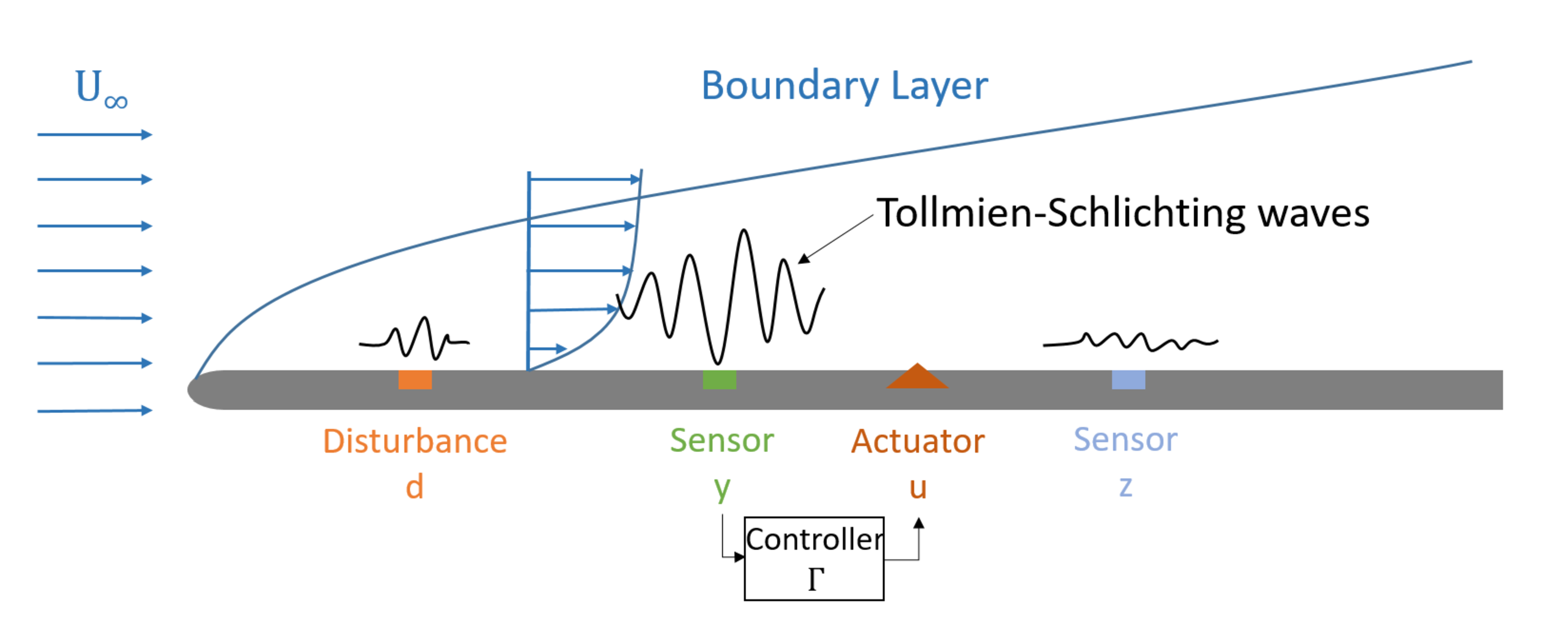}
\caption{Representation of the feed-forward flow control configuration used to attenuate TS waves. Schematic adapted from \cite{Li2021}.}
\label{fig:SchemeCTRL}
\end{figure}

The experimental setup used in this work is similar to the one presented by \cite{Brito2021}, and so it is the control configuration considered to the attenuation of Tollmien-Schlichting waves, which is represented by the schematic shown in Figure \ref{fig:SchemeCTRL}.

Upstream 2D disturbances were inserted in the boundary layer through a slit located at $x/c = 0.1$ in order to trigger the TS waves. The slit was connected to a loudspeaker that provided broadband excitation, as will be described later. In Figure \ref{fig:SchemeCTRL}, such upstream excitation is represented by the disturbance $d$. Electret microphones, represented by $y$ and $z$, were placed tangent to the pexiglass surface to measure pressure fluctuations at $x/c = 0.3$ and $x/c = 0.4$, respectively. The former was used as a reference sensor to determine the actuation $u$ based on the control kernel $\Gamma$, as shown earlier in Eq. (\ref{eq:K_IFFC}). The latter microphone is the control target, which was also used to measure the performance of the controller. The microphones had a signal to noise ratio of 58 dB and an impedance of 2.2 $\mathrm{k\Omega}$, and were powered by means of in-house designed circuits. 

A dielectric barrier discharge (DBD) plasma actuator \cite{Liu1995, Enloe2004} provided the control signal to attenuate the TS waves. Many studies have shown that such a mechanism is suitable for flow control applications \cite{Kissing2021, Zheng2016, Sung2006, Roth}, including for the specific case of boundary layer control \cite{Brito2021, Rodrigues2017, fabbiane2015, Kotsonis2015}. Two copper foil tapes mounted on opposite sides of the pexiglass plate (dielectric material), without overlap, worked as electrodes to ionize the air over the airfoil surface. This induces a near surface forcing, whose intensity can vary accordingly with an input signal, which allows flow control applications.   

The electrodes were connected to a high voltage generator device, Minipuls 2.1 from GBS Elektronik, which provided an electric-potential of 10 kV between them. A Hewlett-Packard frequency generator was used to set the Minipuls at an alternating voltage of about 12 kHz. The plasma generator was powered with 20 V input from a power supply unit manufactured by Minipa, model MPL-3305M.   

Sensor readings and actuation were performed with a DS1103 R\&D Controller Board manufactured by dSPACE. The controller executed these tasks in real time at a sample rate of 10 kHz, based on a Simulink plant model that was converted to the dSPACE ControlDesk environment. Before sending the sensor signals to the controller board, a signal conditioner was used to guarantee an appropriate output voltage level of such signals. The communication between the controller and other devices was performed via BNC connectors. The same type of connectors was used for the communication between the microphones and the signal conditioner.

The perturbations were generated by a 200 W speaker facing a set of 37 flexible tubes of equal length connected to the airfoil slit. The speaker was connected to an 8 $\mathrm{\Omega}$ amplifier. A white noise perturbation was determined via the Simulink plant model, and its amplitude could be set in the dSPACE ControlDesk environment. This information was sent to the amplifier through the dSpace controller board. 

\begin{figure}[!ht]
\includegraphics[width=0.95\textwidth]{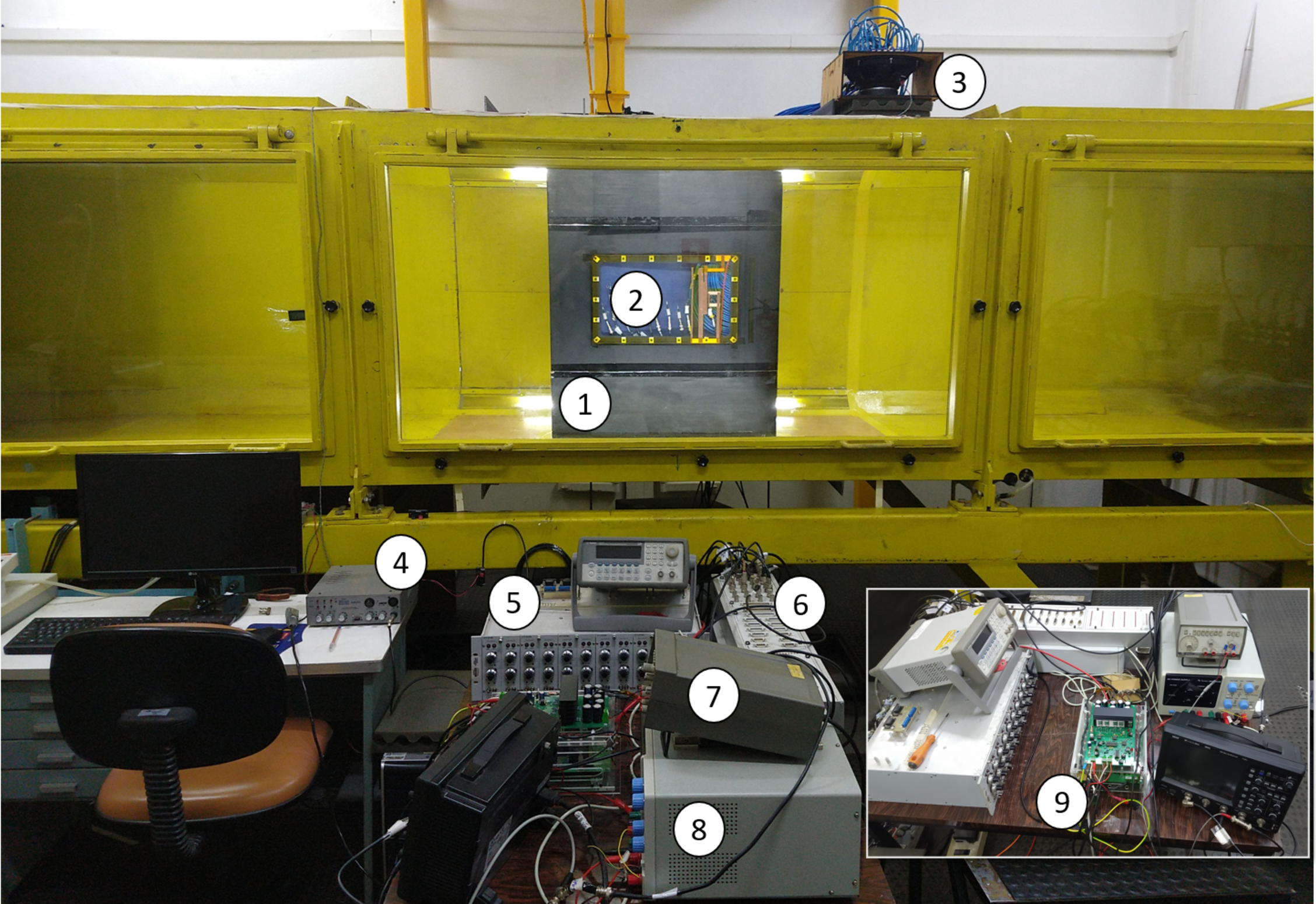}
\caption{Experimental apparatus: 1 - wing profile, 2 - pexiglass plate with the sensors and actuator, 3 - speaker with the flexible tubes, 4 - amplifier, 5 - signal conditioner, 6 - dSPACE acquisition system, 7 - frequency generator, 8 - power supply, and 9 - high voltage generator, Minipuls 2.1.}
\label{fig:Apaexp}
\end{figure}

A general view of the experimental apparatus is shown in Figure \ref{fig:Apaexp}.

\subsection{Preliminary control settings and system identification} \label{PreCtrl}

Before running the experiments to obtain the required data in order to determine the control law and then proceed with the actual boundary layer control, it is necessary to establish the appropriate level of perturbation to trigger linear Tollmien-Schlichting waves, and to determine the level of energy provided by the actuator.

The noise perturbation level determines how early the boundary layer transition will occur and the amplitude of the TS waves when they reach the microphones.  With low perturbation levels, the TS waves are poorly detected by the sensors; on the other hand, high perturbations levels will trigger non-linearities too early. Since the growth dynamics of the Tollmien-Schlichting are linear, the appropriate noise level was determined considering the coherence between the generated disturbance and the sensors $y$ and $z$. The most amplified TS waves under the flow conditions considered here should be observed around the frequency of $200 Hz$ \citet{Brito2021}; thus, high coherence around this region is expected for an appropriate level of the disturbance $d$. It is important to mention that the application of the Wiener-Hopf method does not require the knowledge of $d$ as will be shown in the next sections. However, we have taken advantage of the availability of such information to facilitate some analysis, but we could have considered the coherence between $y$ and $z$, for instance, to observe the amplification of the TS waves. 

For the actuation, we have adjusted a steady plasma intensity that slightly attenuates the TS waves, employing an offset parameter given in volts. This ensures that there is detectable plasma generation, and guarantees at the same time that, when the control is applied, significant attenuation will be mainly due to the control law that modulates plasma amplitudes. The offset parameter ranges from 1.2 V to 1.9 V. The control signals will modulate the plasma amplitude around the offset value determined in the last step, but with their amplitude limited by the offset limits. The offset value used in the experiment was about 1.38 V.

It was also verified if white-noise excitation of the plasma at low amplitudes, of about 0.08 V or less, would lead to a reasonable coherence with respect to the sensor in $z$. This is important to obtain the transfer function between the actuator and the microphone $z$, and also to check if the system responds linearly to small variations of the actuation signal. 

\begin{figure}[h!]
\includegraphics[width=0.95\textwidth]{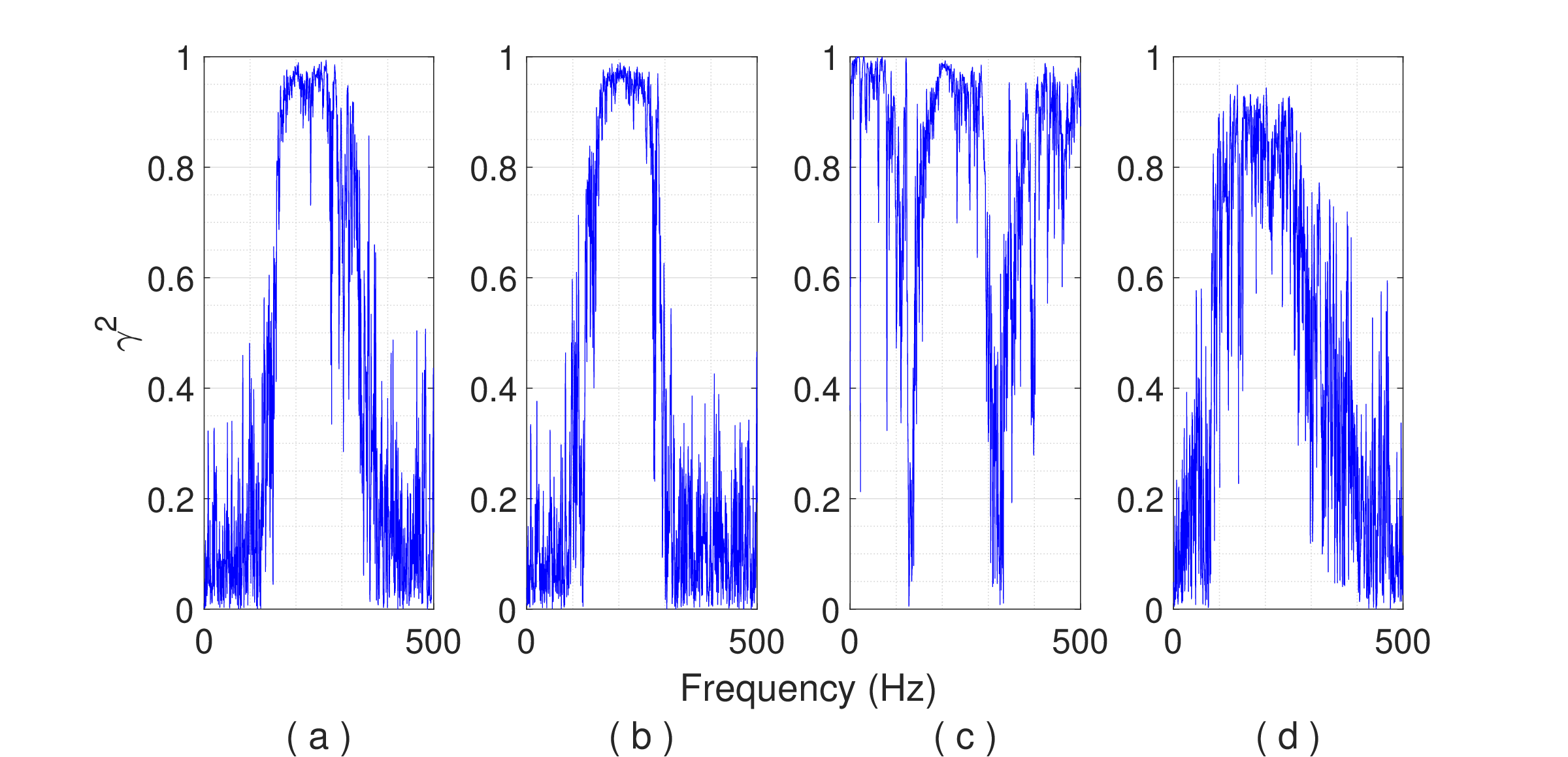}
\caption{Coherence between: (a) perturbation $d$ and reference signal $y$; (b) perturbation $d$ and control target $z$; (c) reference signal $y$ and control target $z$; (c) actuation $u$ and control target $z$.}
\label{fig:Cohe}
\end{figure}

The coherence between the sensors and the actuation and perturbation signals are shown in Figure  \ref{fig:Cohe}.

The plasma offset and the noise perturbation level might need adjustments during the experiments due to weather conditions. The plasma intensity is negatively affected by the environment humidity; thus, the higher the air humidity levels, the higher the offset parameter needs to be. The humidity of the air and temperature/pressure changes will also cause small variations in the Reynolds number; thus, minor changes in the white noise level might need to be applied in order to keep the transition to turbulence around the same region and then preserve the coherence on the same level shown in Figure \ref{fig:Cohe}.  

The experimental data to generate the kernel consisted of $30$ s measurements at a sampling frequency of 10 kHz, where the first 15 s of measurements were used to generated the control kernels, the second half of the data was used to evaluate the performance of the control kernels offline. The PSDs and CSDs necessary to obtain the control kernel, which are discussed in the next section, were obtained via Welch’s method, where a Hamming windowing has been considered with $2^{13}$ sampling points to calculate the discrete Fourier transform, and 50\% overlapping, which yielded 35 blocks.

As will be shown in the next section, the control kernel will be constructed based on three characteristics of the flow control system: the PSD of sensor $y$, $S_{yy} $; the CSD between sensor $y$ and $z$, $S_{yz}$; and the transfer function between $u$ and $z$, defined as $G_{uz} = S_{uz}/S_{uu}$, where $S_{uz}$ is the CSD between the actuation signal $u$ and the control target $z$, and $ S_{uu}$ is the PSD of the actuation signal. The first two characteristics are obtained with the speaker on and the plasma off, while the transfer function $G_{uz}$ is obtained with the speaker off and plasma on, and driven by a white noise signal. 

\begin{figure}[h!]
\includegraphics[width=0.95\textwidth]{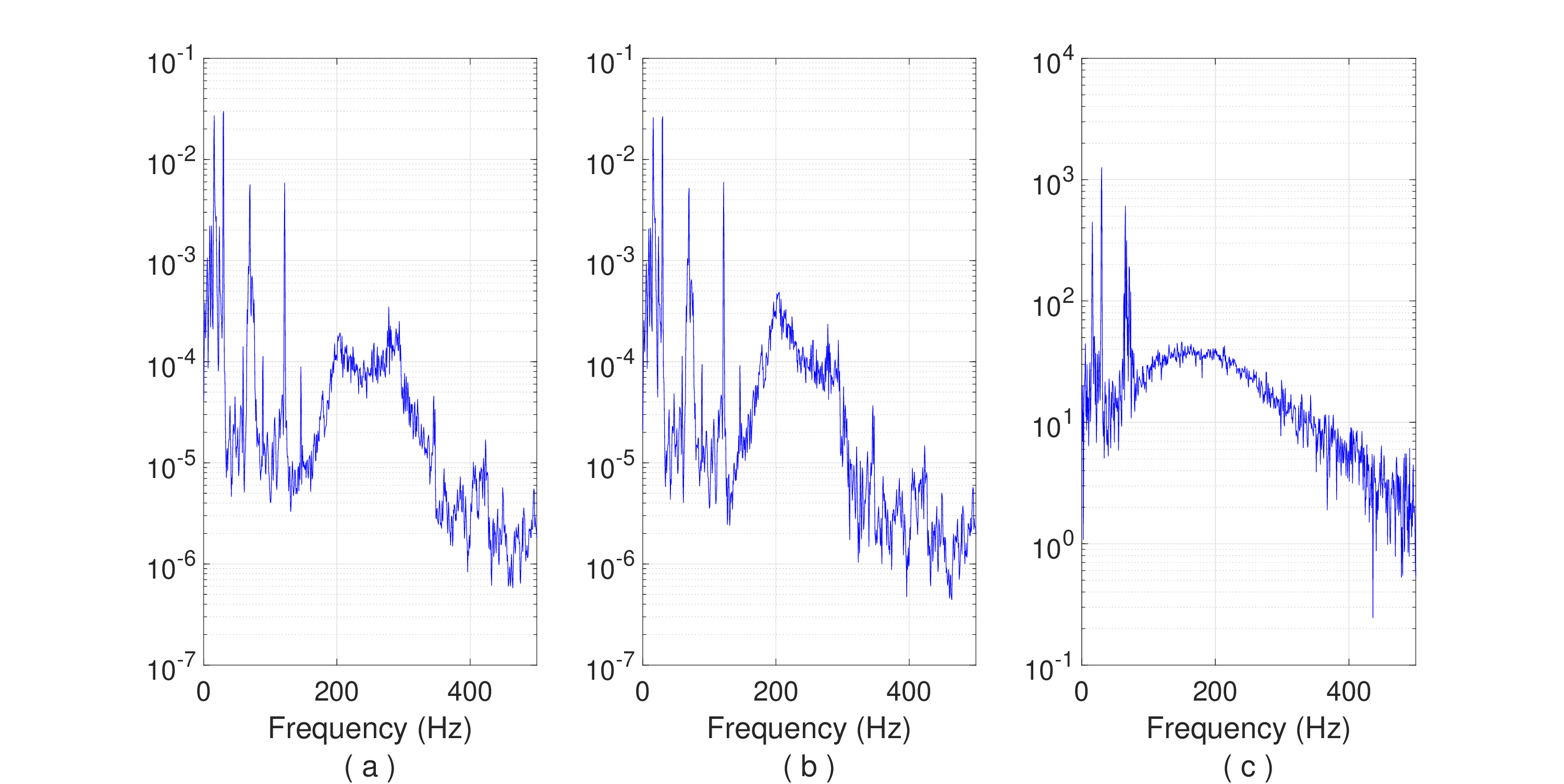}
\caption{ Information extracted from experiments to obtain the control kernel: (a) PSD of measurements in $y$ ($|S_{yy}|$), (b) CSD between measurements in $y$ and $z$ ($|S_{yz}|$), and (c) transfer function between the actuation signal $u$ and measurements in $z$ ($|G_{uz}|$).}
\label{fig:TFs}
\end{figure}

In Figure \ref{fig:TFs}, we show these three functions obtained from the experimental data of the present work. The relevant part of these functions appears for frequencies between 150 Hz and 320 Hz, but, as will be discussed in section \ref{SubUndNoises}, some peaks emerge due to experimental noise. 

Once the parameters discussed above are set up, and the characteristics of the flow control system are obtained, the kernels can be obtained according to the method described in the next section. 

\subsection{The Wiener-Hopf technique applied to control problems}

Optimal control can be obtained by minimizing a quadratic cost functional that balances a deviation cost with a control cost, augmented with Lagrange multipliers $\Lambda_-$ to impose causality ($u(t<0)=0$), as shown in Eq. (\ref{eq:CtrlFunc}). Since we have only one reference sensor ($y$), one actuator ($u$) and one target ($z$), the variables here are scalar (with a frequency dependency). However, the generalization of the problem to the application of multiple sensors and actuators can be easily accomplished \cite{Martini2022}. The size of the matrices to solve the control problem under the Wiener-Hopf approach scales with the number of sensors and actuators. That is an advantage in comparison to the LQR controller, which scales with the size of the system.    

\begin{equation}
\begin{split}
   J & =  \int ^{\infty} _ {-\infty}\left(\langle{{{Q}{z_c^{*}}}{{z_c}}}
  + {Ru_+^{*}}{u_+} \rangle + {{\Gamma_+}{\Lambda} _-} + {{\Gamma^{*}_+}{\Lambda}^{*} _-}\right) dt \\
  & = \int ^{\infty} _ {-\infty} \left(\langle{{{Q}\hat{z}_c^{*}}{\hat{z}_c}}
  + R{\hat{u}_+}^{*}{\hat{u}_+} \rangle+ {{\hat{\Gamma}_+}{\hat{\Lambda}} _-} + {{\hat{\Gamma}_+^{*}}{\hat{\Lambda}}^{*} _-}\right) d\omega
  \end{split}
  \label{eq:CtrlFunc}
\end{equation}

\noindent where the $*$ superscript denotes the complex conjugate, ${z_c}$ is the controlled signal at the target location, ${Q}$ and ${R}$ are the system deviation and control costs, respectively. We also have that ${u=\Gamma * y}$, as defined by Eqs. (\ref{eq:K_IFFC}) and (\ref{eq:kt_IFFC}), where $*$ represents a convolution; and that ${z_c} = {z_{un}} + {G_{uz}}*{u} = {z_{un}} + {G_{uz}}*{\Gamma*y}$, i.e., the output is  a linear combination of the uncontrolled signal and the actuation response, where $z_{un}$ is the uncontrolled signal at the target location. The subscripts + and - indicate that the functions are regular in the upper and lower half-plane of the complex $\omega$ plane, respectively. When transformed to time domain, the plus functions are zero for $t > 0$, while the minus functions are zero for $t < 0$. 

Expanding Eq. (\ref{eq:CtrlFunc}) we find, 

\begin{equation}
 \resizebox{\textwidth}{!} {$
\begin{split}
 J & =  \int \left(\langle{{{Q}(\hat{z}_{un} + \hat{G}_{uz}\hat{\Gamma}_+\hat{y})}^{*}{(\hat{z}_{un} + \hat{G}_{uz}\hat{\Gamma}_+\hat{y})}}
  + R{(\hat{\Gamma}_+\hat{y})}^{*}{(\hat{\Gamma}_+\hat{y})} \rangle+ {{\hat{\Gamma}_+}{\hat{\Lambda}} _-} + {{\hat{\Gamma}_+}^{*}{\hat{\Lambda}}^{*} _-}\right) d\omega \\
&   =  \int \left({Q}({{S_{zz} + \hat{G}_{uz}\hat{\Gamma}_+S_{zy} + \hat{G}^{*}_{uz}\hat{\Gamma}^{*}_+S_{yz}  + \hat{G}^{*}_{uz}\hat{G}_{uz}\hat{\Gamma}^{*}_+\hat{\Gamma}_+S_{yy}}}) + {R\hat{\Gamma}_+}{\hat{\Gamma}^{*}_+S_{yy}} + {{\hat{\Gamma}_+}{\hat{\Lambda}} _-} + {{\hat{\Gamma}_+}^{*}{\hat{\Lambda}}^{*} _-}\right) d\omega 
 \end{split}
$ }
  \label{eq:CtrlFuncExp}
\end{equation}

 Minimizing the quadratic cost functional with respect to ${\hat{\Gamma}^{*}_+}$, one obtains a Wiener-Hopf equation \cite{Martini2022}, which, written in terms of quantities that may be obtained directly from experiments, is given by

 \begin{equation}
    {{\hat{G}^{*}_{uz}QS_{yz}  + (\hat{G}^{*}_{uz}Q\hat{G}_{uz} + R)\hat{\Gamma}_+S_{yy}}} + {\hat{\Lambda}}^{*} _- = 0
  \label{eq:WHAnt}
\end{equation}

\noindent we can still define ${\hat{H}_l}  = {\hat{G}_{uz}}^{*}{Q}{\hat{G}_{uz}}+{R}$ and ${\hat{H}_r} = -{\hat{G}_{uz}}^{*} {Q}$, and rewrite the Wiener-Hopf equation for the control problem as

\begin{equation}
 {\hat{H}_l}{\hat{\Gamma}}_+{S_{yy}}+{\hat{\Lambda}}^*_{-} = {\hat{H}_r}{S_{yz}},
  \label{eq:CtrlWH}
\end{equation}

\noindent where ${S_{yy}}$ is the PSD of measurements from sensor $y$, and ${{S_{yz}}}$ is the cross-spectral density (CSD) between signals in $y$ and $z$ in open loop. Comparing to the formulation given by \citet{Martini2022}, we note that $S_{yy}$  is equivalent to $\hat{G}_l$ and $S_{yz}$  to $\hat{G}_r$.  Although this has been discussed in \citet{Martini2022}, the above derivation does not assume a linear time-invariant system, relying only on the linear effect of the actuator on the target $z$.  

The Wiener-Hopf control equation, Eq. (\ref{eq:CtrlWH}), can be solved using additive and multiplicative factorization, as exemplified at the Appendix, which yields the optimal causal control kernel given by 

\begin{equation}
    {\hat{\Gamma}}_+ = {\hat{H_l}_+^{-1}}( {\hat{H_l}_-^{-1}} {\hat{H_r}} {{S}_{yz}} {{S}^{-1}_{yy-}})_+ {{S}_{yy+}^{-1}}
    \label{eq:Gammaplus}
\end{equation}

For analytical and numerical details on this solution we refer the reader to \citet{Martini2022}. 

When the inverse Fourier transform is applied to ${\hat{\Gamma}}_+$, one obtains ${{\Gamma_+(\tau<0)}}  = 0$. Thus, the actuation signal, given by Eq. (\ref{eq:kt_IFFC}), does not depend on future sensor readings in $y$, and the kernel is causal by construction. Furthermore, from the CSDs, the forcing color, which has been shown to be important by previous studies\cite{martini2020,amaral2021}, is implicitly considered in the Wiener-Hopf control approach. 

$S_{yy}$, $S_{yz}$ and $G_{uz}$, which are the three characteristics necessary to obtain an optimal causal solution in the frequency domain, are shown in Figure \ref{fig:TFs}, section \ref{PreCtrl}.

The inverse feed-forward control method, which is used in this work as a comparison, can be obtained from Eq. (\ref{eq:CtrlWH}) by simply dropping the Lagrange multipliers and ignoring the factorization.

\begin{figure}[h!]
\includegraphics[width=0.95\textwidth]{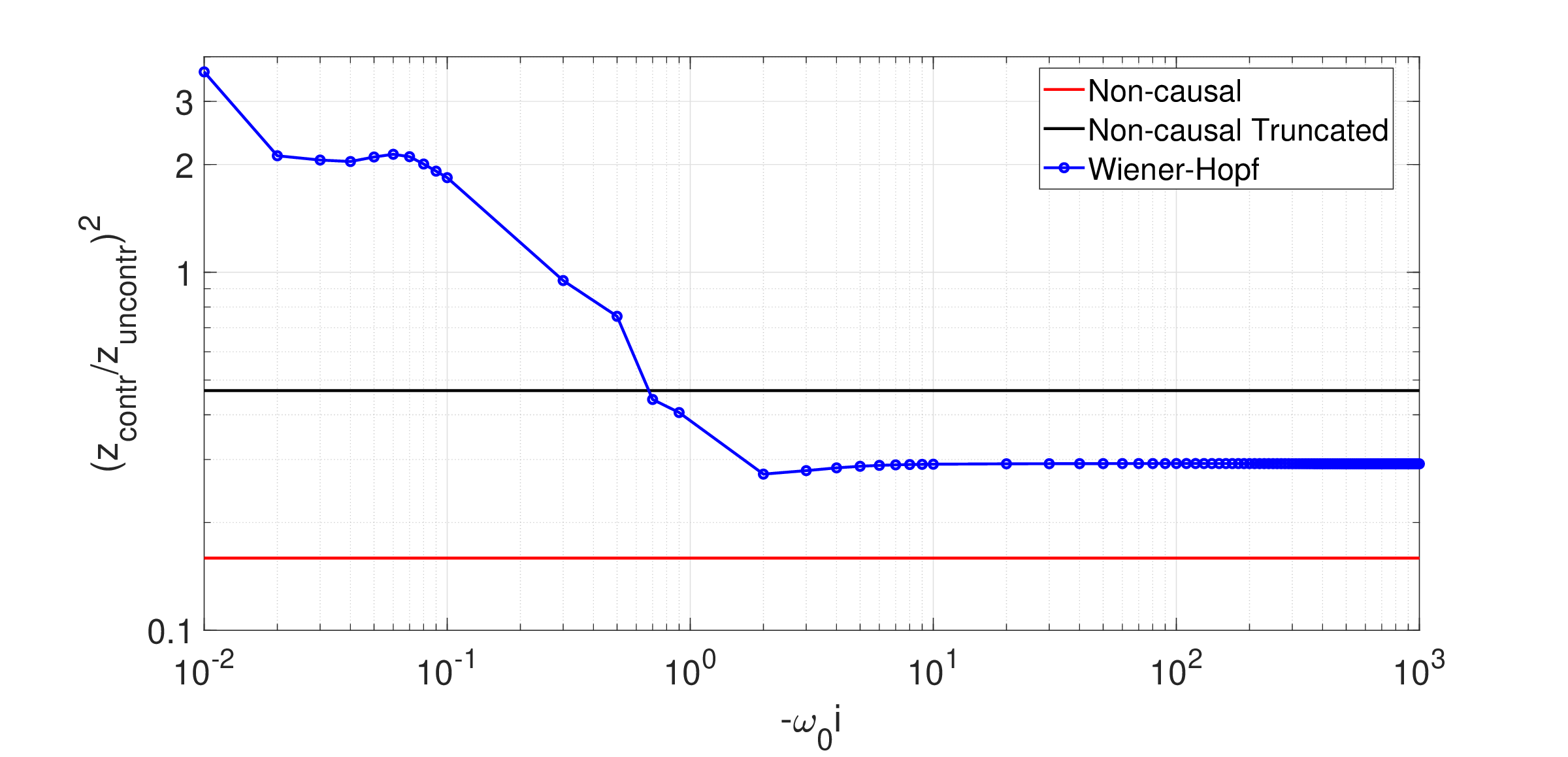}
\caption{Convergence of the control kernel with $\omega_0$}
\label{fig:OmegaVar}
\end{figure}

In order to perform the multiplicative factorization, an arbitrary frequency parameter $\omega_0$ has to be chosen \cite{daniele2007,Martini2022}. \citet{daniele2007} mention that $\omega_0$ introduces an apparent singularity that might increase the numerical instability when the factorization is performed numerically, and, in order to avoid this issue, suggest choosing $\omega_0$ for the region of a singularity related to the physical problem. In our case, the specific region of singularities is not obvious, however, since we are interested in functions that are regular in the upper-half plane, any singularity should appear in the lower-half plane, i.e., $\omega_0$ has to be a negative imaginary number. We have considered a purely imaginary number and checked the convergence of the kernel's performance with respect to different values of $\omega_0$. The performance was predicted from an offline simulation for a fixed value of $R$, the result of the convergence analysis is shown in Figure  \ref{fig:OmegaVar}. We have chosen $\omega_0 = -50i$ for the multiplicative factorization, the solution is sufficiently converged. 

\subsection{Dealing with undesired noises captured by the sensors} \label{SubUndNoises}

From the spectra shown in Figure \ref{fig:TFs}, one may notice that the signals are dominated by some low frequency noise, in the range of 17 Hz to 121 Hz, whereas the region of interest is from 150 Hz to 320 Hz, which are related to the excitation of Tollmien-Schlichting waves and, thus, the region of higher coherence between the sensors, as shown in Figure \ref{fig:Cohe}. The wind tunnel fan was identified as the source of the 16 Hz, noise and the 30 Hz peak was identified as an open-pipe resonance of the wind tunnel \cite{Brito2021}. The peaks at 70 Hz and 121 Hz appear to be related to electromagnetic interference, and are not always present. The microphones used as pressure sensors capture the fluctuations related to the TS waves and also acoustic waves in the test section. 

The simplicity and low cost corresponding to such microphones make them a compelling option for flow control applications. However, it is important to mitigate the effect of other noise sources, not related to the TS waves that should be aimed by the controller.     
Otherwise, such low frequency noise substantially increases the amplitudes of the actuation signal, which is detrimental to the controller performance \cite{Brito2021}. The Wiener-Hopf technique, besides yielding a causal solution, provides tools to mitigate the effect of the undesired noise on the controller. Since earlier studies have demonstrated a high relevance of such aspect \cite{Brito2021}, this issue has been addressed by this present work.

\begin{figure}[!ht]
\includegraphics[width=0.95\textwidth]{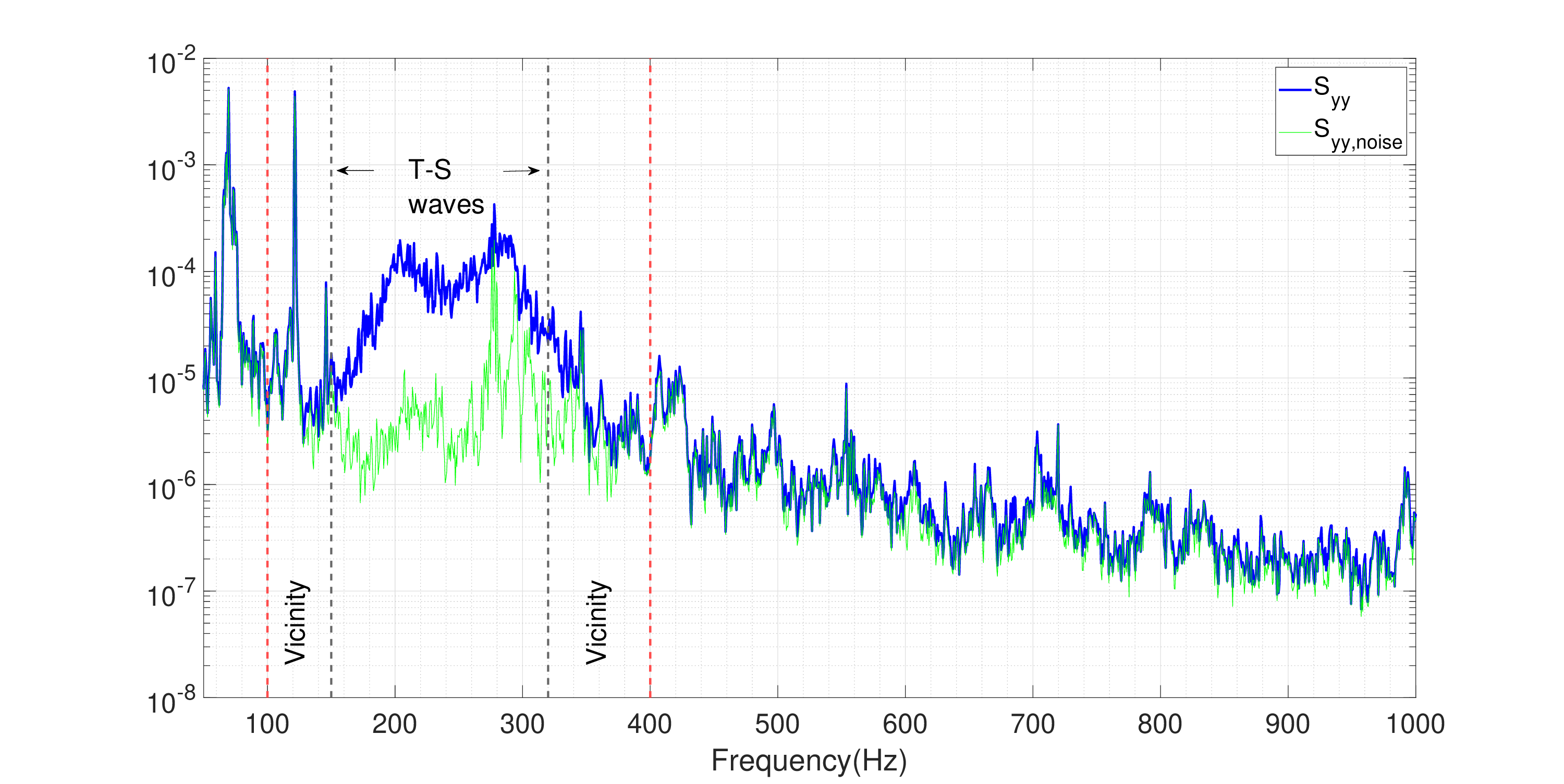}
\caption{Spectrum of the measured signal in $y$, ${S_{yy}}$, and its estimated noise, ${S_{yy_{noise}}}$}
\label{fig:Noise}
\end{figure}

To mitigate the effect of noise not related to fluctuation of TS waves, it was considered the sum of the signal in $y$ with its estimated noise, defined as 

\begin{equation}
{{S_{yy_{modified}}}}(\omega) = {{S_{yy}}}(\omega) + M{S_{yy_{noise}}}(\omega)
\label{eq:Syymod}
\end{equation}

\noindent where the factor $M$ determines how much of the frequency content appears as noise in $y$, affecting ${S_{yy}}$ (power spectral density of the $y$ sensor) but not ${S_{yz}}$ (cross-spectral density between y and $z$). $M=50$ has been considered to sufficiently reduce the undesired noise effect on the controller. We consider as noise the part of the sensor signal which cannot be estimated by the disturbance signal $d$. The estimated noise ${S_{yy_{noise}}}$ is defined by Eq. (\ref{eq:est_noise}) \cite{bendat2010random}, and it is shown in Figure \ref{fig:Noise}.

\begin{equation}
{S_{yy_{noise}}}= 
\begin{cases}
     0,& \text{if } 100 \leq \mathrm{F (Hz)}\leq 400 \\
    {S_{yy}} - |{{S_{dy}}}|^2/ {{S_{dd}}},              & {\mathrm{otherwise}}
\end{cases}
\label{eq:est_noise}
\end{equation}

\noindent where ${{S_{dy}}}$ is the CSD between the disturbance $d$ and sensor $y$, and ${{S_{dd}}}$ is the PSD of $d$. 

This approach basically applies a control penalisation as a function of the frequency and the respective signal energy level. Thus, for the frequency range related to the excitation of TS waves, and its vicinity, this signal treatment is not applied, i.e., ${S_{yy_{noise}}}$ is defined as 0 for frequencies between 100 and 400 Hz. Otherwise, it would cause a drop in the performance of the controller. Considering the produced used to obtain ${S_{yy_{noise}}}$, we may not be removing some of the acoustic resonances in the tunnel, if those are driven by $d$; however, we observed that these are mainly driven by external factors, not being correlated to $d$, they are, thus, filtered out by this approach. 

\begin{figure}[!ht]
\includegraphics[width=0.95\textwidth]{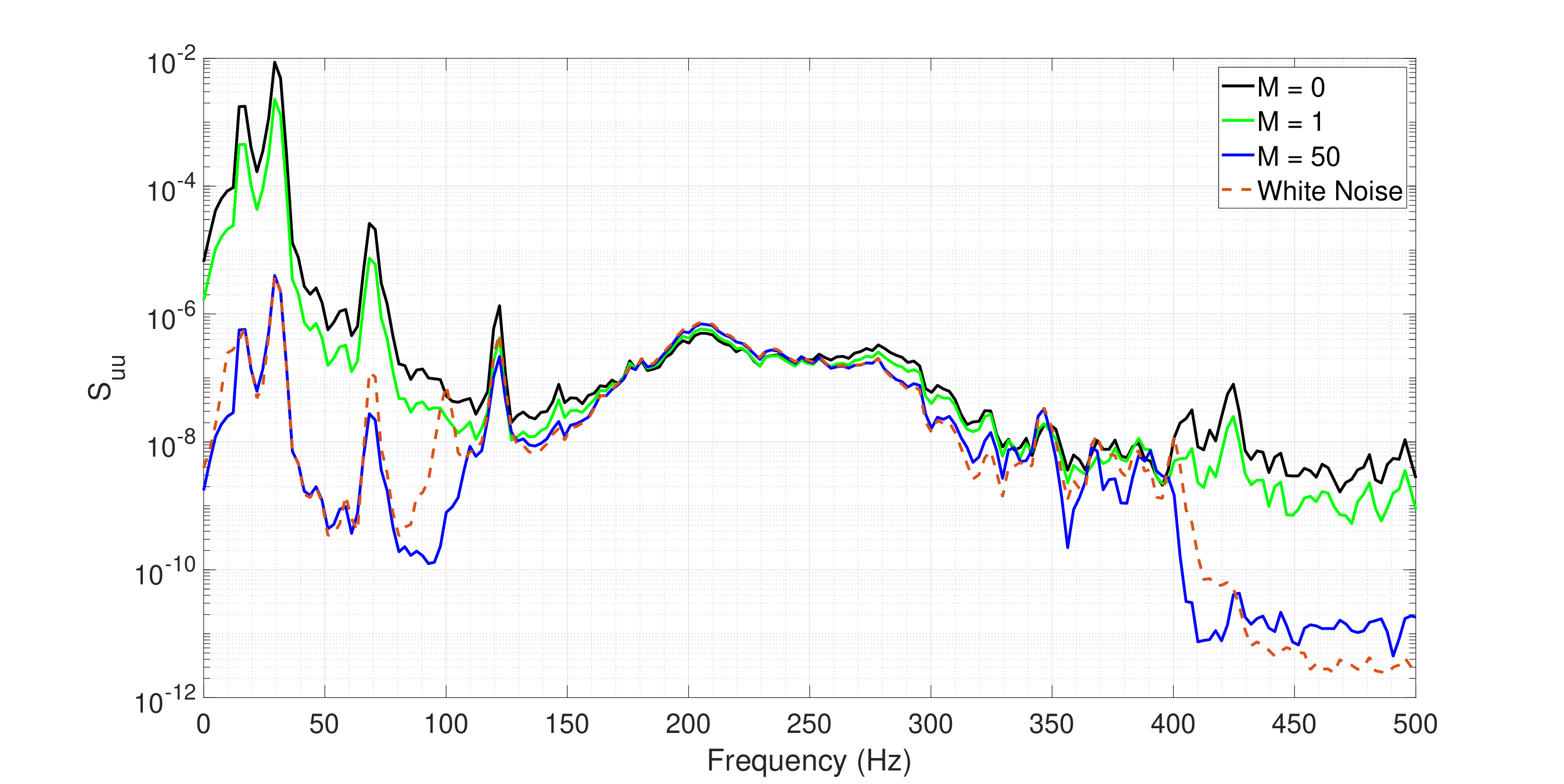}
\caption{\label{fig:Suuex}Spectrum of the predicted actuation signal, $S_{uu}$, considering a noise attenuation approach based on the sensor noise.}
\end{figure}

Figure \ref{fig:Suuex} shows the impact of such signal treatment on the actuation signal, which was predicted from an offline simulation, based on the same data used to obtain the kernels. From Figure \ref{fig:Suuex}, we observed that by simply adding the noise to the sensor once, the noise is barely attenuated; however, by doing it multiple times, sufficient attenuation can be reached. This is important because the low-frequencies noise leads to amplitudes that would saturate the actuator signal, which compromises the controller's effectiveness. By using $M=50$, most of the noise unrelated to the TS waves is reduced by approximately two orders of magnitude when compared to the $M=0$ case, with negligible impact for the frequencies related to the TS waves. However, this signal treatment was only applied to the Wiener-Hopf approach due to the greater sensitivity to these external noises, as matrix factorizations imply that different frequencies are coupled in the solution. We emphasize that the application of Wiener-Hopf approach doesn't rely on the knowledge of the perturbation. Alternatively, the second expression shown in Eq. \ref{eq:est_noise}, could be substituted by an appropriate level of white noise signal. Here it was observed that, $S_{yy_{noise}}$ could be equal to $25 \%$ of the maximum value of $S_{yy} $, out of the region of interest, to attenuate the undesired noises captured by the sensors, as shown in Figure \ref{fig:Suuex}. 

\section{\label{sec:results} Results and discussion}

\begin{figure}[!ht]
\includegraphics[width=0.95\textwidth]{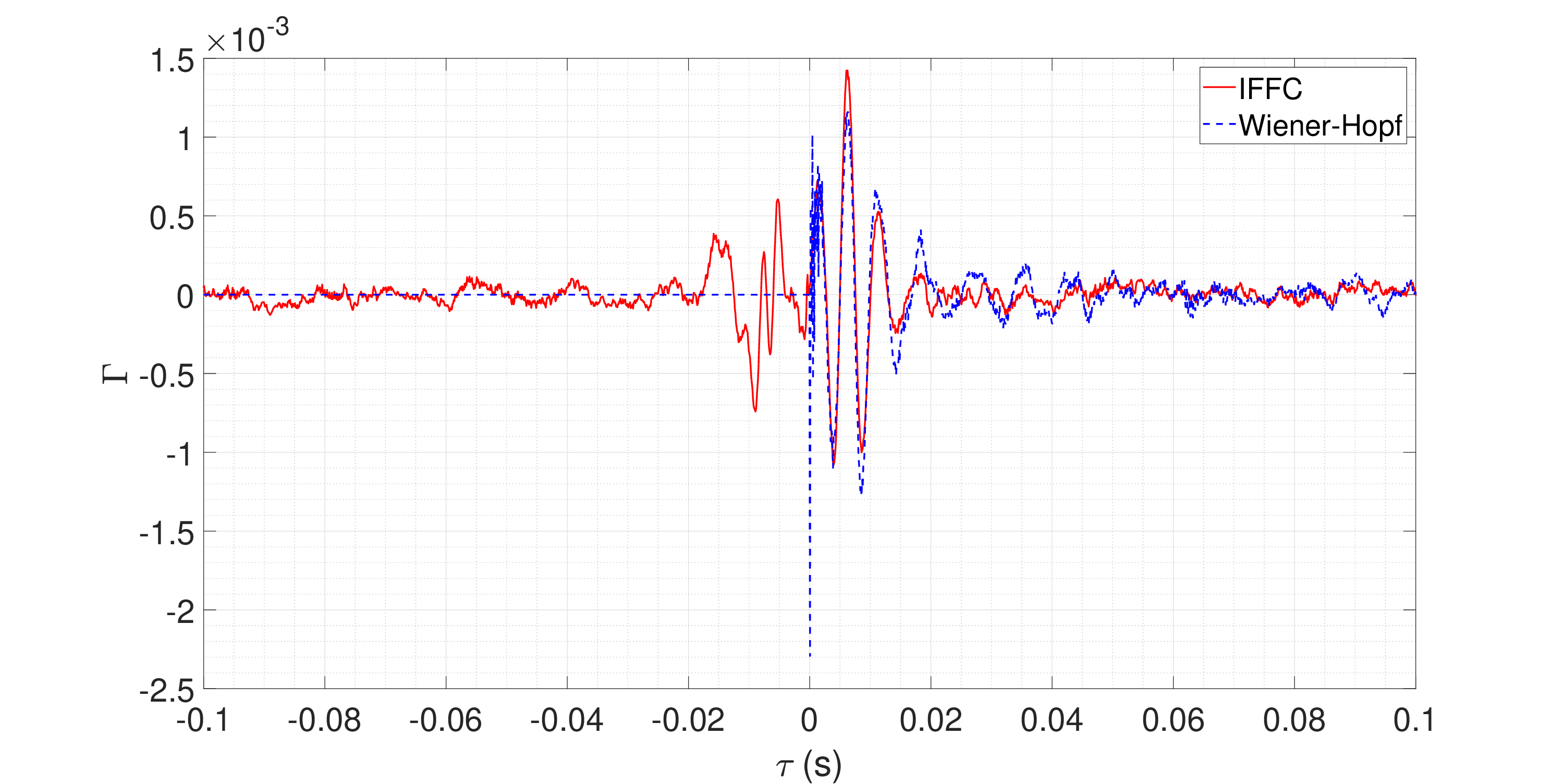}
\caption{\label{fig:kernels_NoDelay}Kernels obtained from the IFFC method and the Wiener-Hopf technique for $R = 300$.}
\end{figure}

As mentioned earlier, the control kernel may be obtained with the functions shown in Figure \ref{fig:TFs}. In what follows we specify $Q=1$ and study the effect of control penalisation $R$. As an example, in Figure \ref{fig:kernels_NoDelay} is shown the kernels obtained for the specific case of a control penalisation $R = 300$, where the Wiener-Hopf kernel is compared with the non-causal solution obtained with the IFFC approach. Here we compare the non-causal IFFC and the Wiener-Hopf kernels, so the anti-causal part, which is discarded in IFFC, be evident.

The results in Figure \ref{fig:kernels_NoDelay} show that the anti-causal part present in the IFFC kernel is partially compensated by the peak at $\tau\approx 0$ in the Wiener-Hopf solution, as discussed in \citet{Martini2022}. A similar observation was made by comparing LQG and IFFC controllers in \citet{morra2020}. Different values of control penalisation were considered for the control of Tollmien-Schlichtig waves, where the performance of these kernels was obtained considering the ratio of RMS values of the $z$ signal, band-pass filtered in the frequency range of interest: 150 to 320 Hz, defined as 

\begin{equation}
    Perf_{CTRL} =  \frac{rms( z_{controlled})}{rms(z_{uncontrolled})}
    \label{eq:P_ctrl}
\end{equation}

\begin{figure}[!ht]
\includegraphics[width=0.95\textwidth]{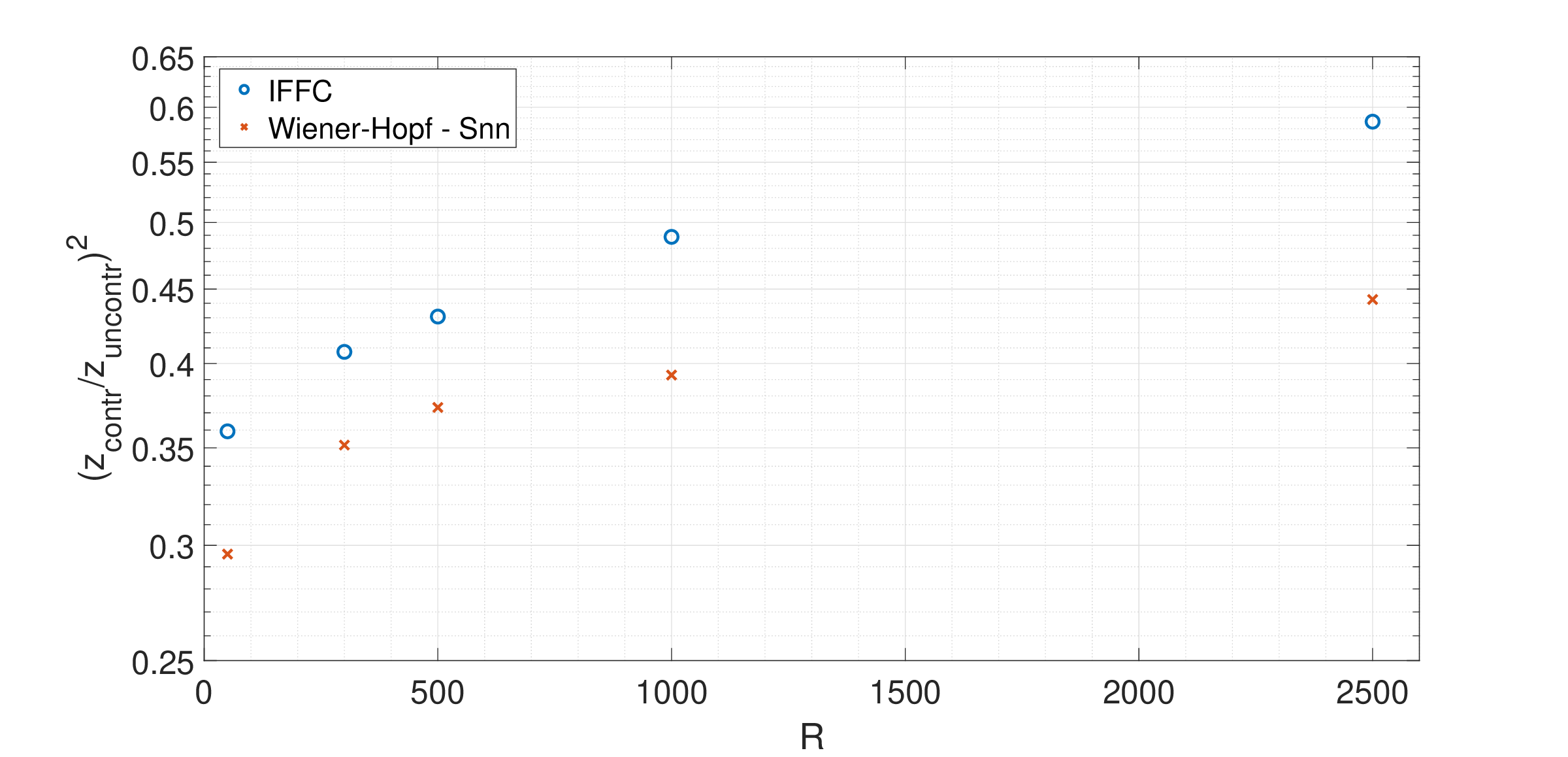}
\caption{\label{fig:Perf_5Rs_Nodelay} Average performance of the kernels obtained at the experimental control of TS waves for different values of control penalisation, $R$.}
\end{figure}

\begin{figure}[!ht]
\includegraphics[width=0.95\textwidth]{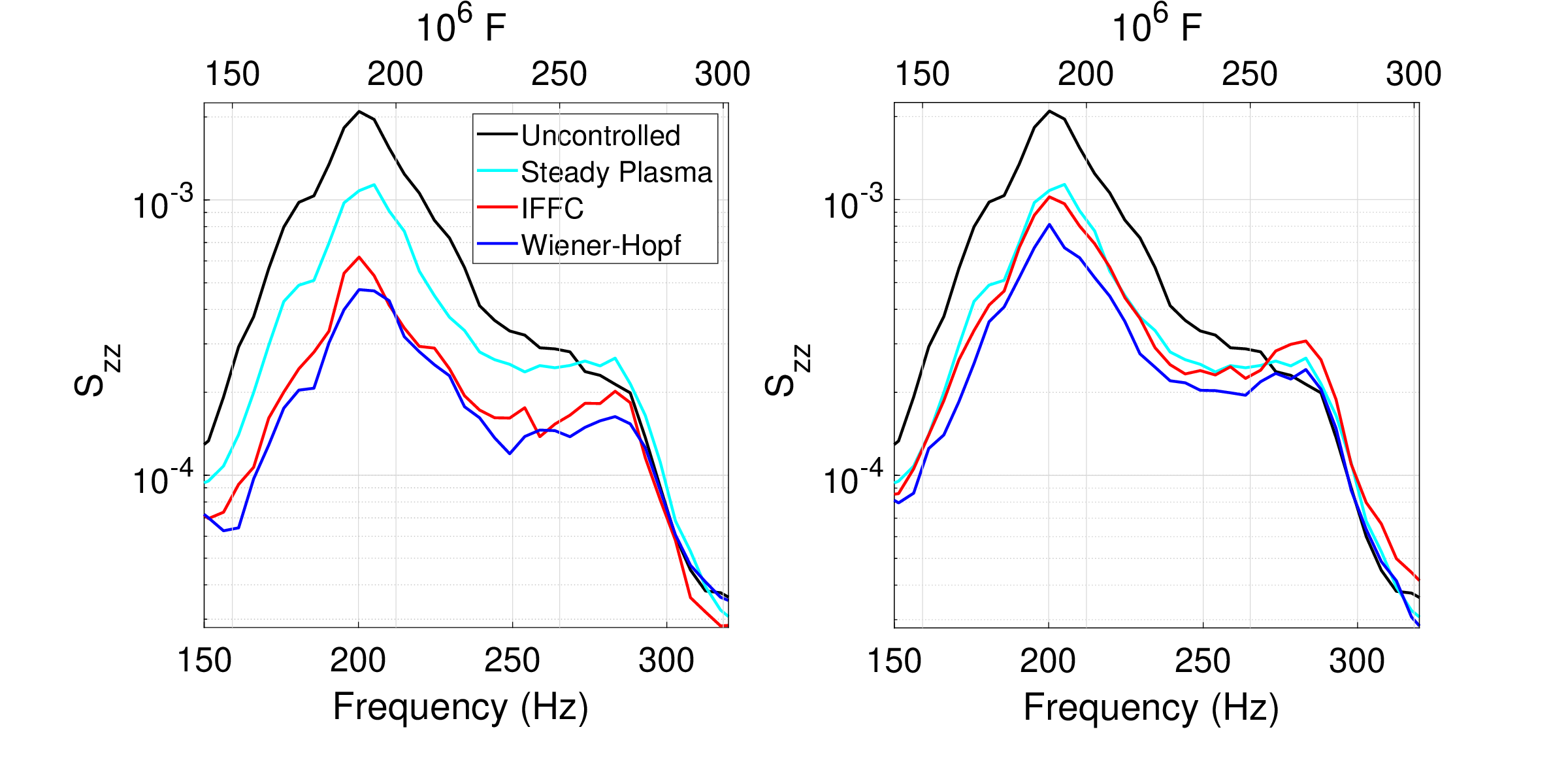}
\caption{\label{fig:Perf_Rbest_NoDelay}Comparison of the uncontrolled and controlled signals in $z$ for R = 50 (left), and R = 2500 (right). The top axis shows the non-dimensional frequency $F = (2\pi \nu/U_{\infty}^2)f$.}
\end{figure}

The various control setups considered in this work were sampled in an alternated manner, with IFFC results followed by the Wiener-Hopf control applied with the same conditions. In total, $5$ measurements, of $5 s$ each, were performed for each case. The performance results are shown in Figure \ref{fig:Perf_5Rs_Nodelay}. 
The Wiener-Hopf solution yielded better results than the truncated IFFC method for a same value of R. In Figure \ref{fig:Perf_Rbest_NoDelay}, the spectrum of the signal in $z$ is shown considering the best result obtained with each method, where is also shown the uncontrolled signal and the signal obtained with an open-loop control actuation (steady plasma) for comparison.

The spectra obtained with the two controllers are relatively similar, with lower PSDs obtained with the Wiener-Hopf kernel, leading to a relative difference in the performance of about $20 \%$ for $R = 50$, which is the one that yielded the best performance in both approaches. This benefit is in agreement with the optimality property of the Wiener-Hopf kernel. It is expected that this difference in performance becomes more substantial in situations where the anti-causal part of the IFFC kernel becomes more relevant. In order to verify such a condition without modifying the experimental setup, an artificial delay was imposed in the actuation, which emulates a condition where the sensor and the actuator are closer to each other. This forces the actuator to act with less information about the incoming disturbances. Although potentially reducing the accuracy of the estimation, and thus of the control, bringing the actuators closer to the sensors can allow the actuator to act on the TS waves before they are further amplified, thus reducing the actuator energy costs. This was done by modifying the actuation signal $u$ that is used to obtain the transfer function $G_{uz}$. A time delay was applied to signal $u$ before the CSDs were computed. This procedure avoids the need to change the configuration of the experiment, i.e., modifying the actuator's location. Furthermore, no additional data is required to obtain the new kernels. 

\begin{figure}[!ht]
\includegraphics[width=0.95\textwidth]{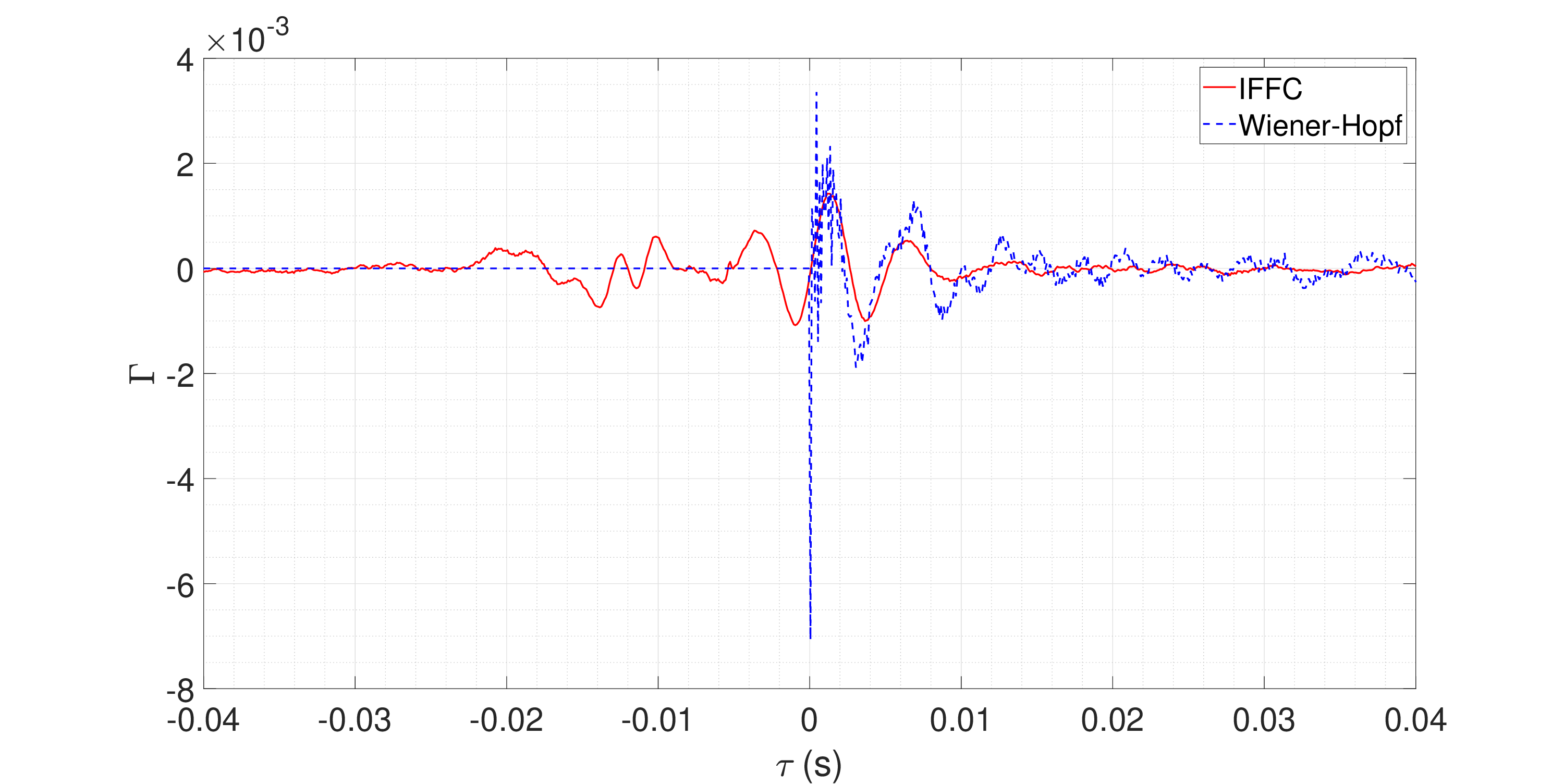}
\caption{\label{fig:Kernels_delay50} Kernels obtained from the IFFC method and the Wiener-Hopf technique for $R = 300$, considering a delay of $4.9$ ms.}
\end{figure}

The kernel considering a delay of  $4.9$ ms, emulating the displacement of the actuator $20$ mm upstream, is shown in Figure \ref{fig:Kernels_delay50} for $R=300$.

\begin{figure}[!ht]
\includegraphics[width=0.95\textwidth]{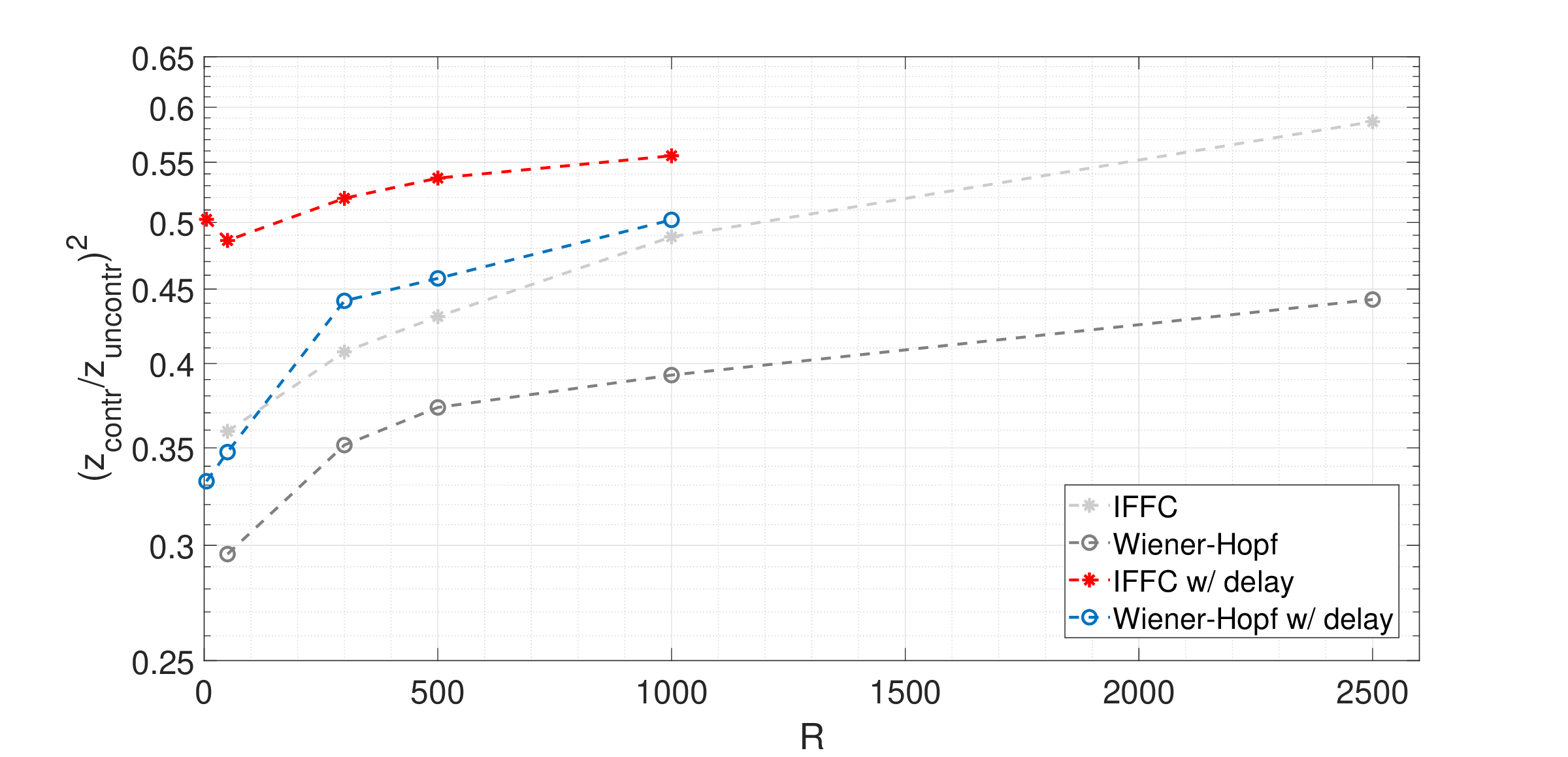}
\caption{\label{fig:Perf_5Rs_delay50} Average control performance obtained in experiments with the IFFC and the Wiener-Hopf techniques when an artificial delay is imposed in the actuation, compared with the results shown in Fig. \ref{fig:Perf_5Rs_Nodelay}.}
\end{figure}

\noindent Here can be noticed that the anti-causal part of the IFFC kernel becomes more significant; in turn, a more prominent peak is observed in the Wiener-Hopf kernel compared with the condition without delay, shown in Figure \ref{fig:kernels_NoDelay}. The performance of the controllers considering the truncated solution of the IFFC and the Wiener-Hopf approaches is shown in Figure \ref{fig:Perf_5Rs_delay50} for different values of $R$. Here it was not considered the case of $R = 2500$, because this condition would not lead to a substantial attenuation of the TS waves. On the other hand, it was possible to obtain a better result for the Wiener-Hopf kernel by using $R = 5$.

The truncated solution obtained with the IFFC method is substantially affected by the increased non-causality of the optimal kernel; its truncation to the causal part compromises performance. On the other hand, the drop in performance of the Wiener-Hopf approach due to the artificial delay is not so substantial if we consider an optimal control penalisation. This was expected since the non-causality becomes more relevant in the IFFC kernel, and simply truncating the non-causal solution will ignore a more significant part of the kernel. With the optimal causal solution provided by the Wiener-Hopf kernel, it is possible to obtain a performance better than the IFFC even considering a delay in the Wiener-Hopf kernel but not in the IFFC one.

\begin{figure}[!ht]
\includegraphics[width=0.95\textwidth]{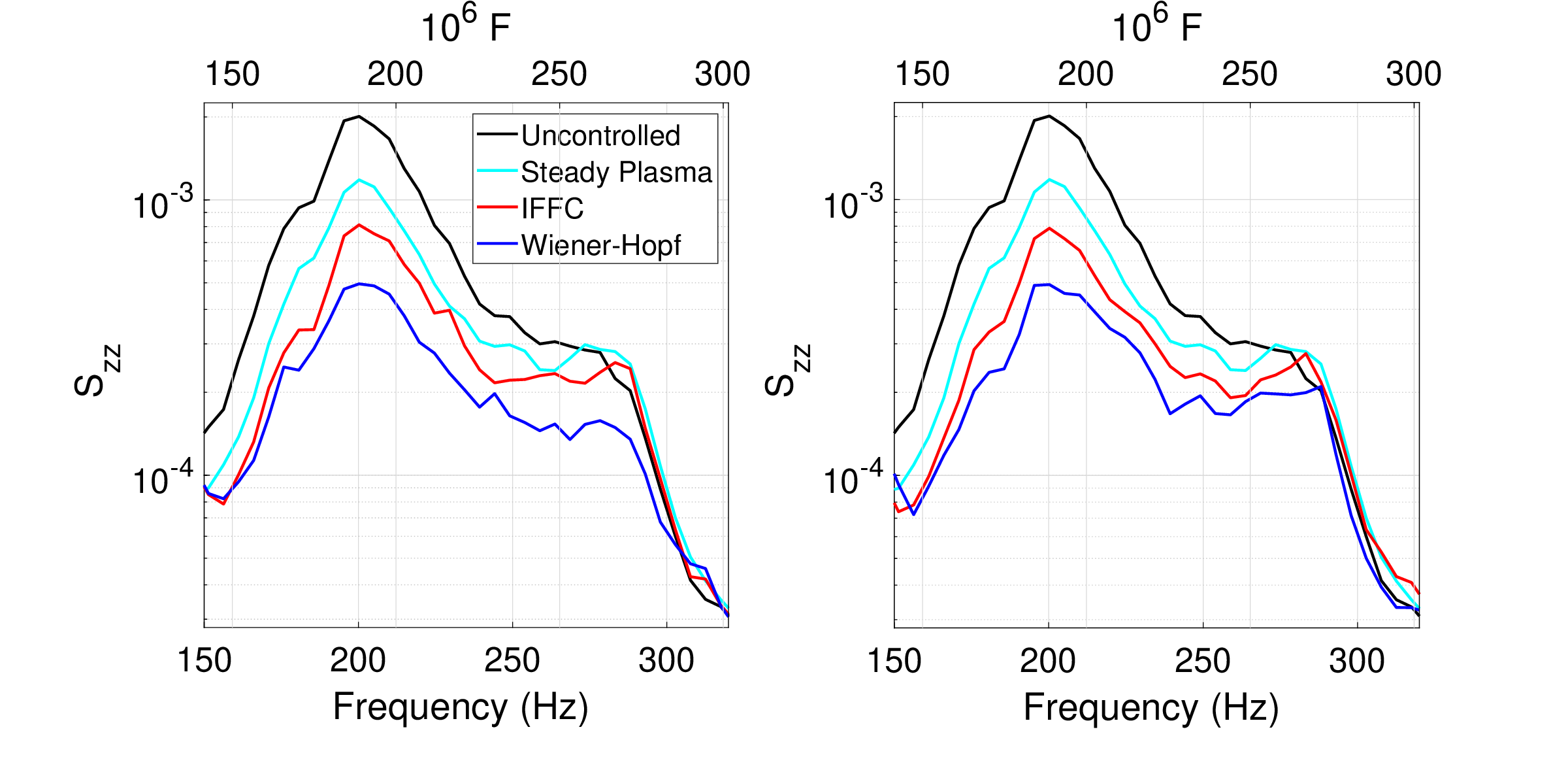}
\caption{\label{fig:Perf_Rbest_Delay50}Comparison of the uncontrolled and controlled signals in $z$ when a delay is artificially applied in the actuation for R = 5 (left), and R = 50 (right). The top axis shows the non-dimensional frequency $F = (2\pi \nu/U_{\infty}^2)f$.}
\end{figure}

The spectrum of the readings in $z$ (control target) is shown in Figure \ref{fig:Perf_Rbest_Delay50}. The uncontrolled signal and the open loop steady plasma is also shown for comparison. With the artificial delay, a greater difference in performance is observed between the IFFC and the Wiener-Hopf, this time corresponding to a relative difference of 31.25 \%, considering $R = 5$ for the Wiener-Hopf and $R=50$ for the IFFC, which corresponds to the control penalisation that yielded the best performance in each case.

\section{\label{sec:conclusion} conclusion}

The use of the Wiener-Hopf method for experimental control of Tollmien-Schlichting waves around a NACA 0008 airfoil at zero incidence has been explored in this manuscript. To the best of our knowledge, this is the first time such a method has been used experimentally in flow control. The Wiener-Hopf method is an approach for optimal control, which is obtained in the frequency domain with the introduction of Lagrange multipliers to ensure that the resulting kernel is causal, i.e., only past sensor information is used to decide control action. In addition, the approach was compared against the inverse feed-forward control method used in previous works \cite{sasaki2018,Brito2021}.

The better performance of a Wiener-Hopf controller compared with a truncated solution is expected from a theoretical point of view. However, the implementation of such a method for flow control has not been much explored so far. The favorable results obtained \modif{previously, in numerical simulation studies \cite{Martinelli2009, Martinelli2009paper, Martini2022}}, motivated the use of the Wiener-Hopf method in an experimental flow control application, which was explored in this work. In particular, we \modif{showed} how Wiener-Hopf kernels may be built directly from spectra and transfer functions obtained from experiments, in a data-driven approach that does not need the construction of reduced-order models. The use of experimental data poses new challenges, and some observed with this work include the limitations in amplitude that the actuator can reach without saturation, and the sensitivity of the microphones to external noises. A noise attenuation approach based on the noise captured by the sensors could circumvent the issues intrinsic to the experimental apparatus considered in this work.

The Wiener-Hopf approach was effectively used for the control of Tollmien-Schlichting waves, yielding better results than the truncated solution of the inverse feed-forward control method. Considering the experimental configuration and characteristics of previous works \cite{Brito2021}, the Wiener-Hopf approach could provide a performance improvement of 20 \% compared to the IFFC technique. In addition, we showed that other configurations of the experiments could increase this difference, as exemplified by the inclusion of an additional actuator delay. As the non-causal aspects of the IFFC kernel becomes more relevant, it is foreseen that the performance of a Wiener-Hopf based controller is less affected than the truncated IFFC kernel. Such behavior of these methods could be observed experimentally with the introduction of an artificial delay applied to the controllers, which corresponds to moving the actuator upstream. This happens because the Wiener-Hopf approach takes into account the causality requirement of the kernel in its formulation, which results in an optimal solution. On the other hand, in the IFFC method, where the kernel is obtained in the frequency domain without any causality requirement; a non-causal kernel is often obtained, and the non-causal part needs to be truncated, leading to performance loss. 

The present work has demonstrated, for an experimental flow control problem, the possibility of getting better results than a typical wave cancellation approach while still obtaining the controller in the frequency domain, using directly experimental power and cross spectra obtained experimentally, without resorting to a reduced-order model. This is a promising approach for flow control, which may enable other applications in wind tunnel or flight.
 
\begin{acknowledgments}
We wish to acknowledge the financial support from the Coordination for the Improvement of Higher Education Personnel (CAPES) and from the São Paulo Research Foundation (FAPESP), grant 2019/26546-6. The airfoil and the high voltage generator used in the experiments were provided by SAAB, obtained through VINNOVA Projects PreLaFlowDes II (2019-00079) and SWE-DEMO (2015-06057). We also acknowledge the support provided by the technicians from the Kwei Lien Feng laboratory at ITA, in special, Wilson Mendes de Souza, who assisted the experiments conducted for this work, which was of great importance. We also would like to acknowledge Marilda V. L. Gontijo, who made it possible to extend the daily use of the laboratory facilities, which was essential to run the necessary experimental cases.  
\end{acknowledgments}

\appendix*
\section{An overview on the Wiener-Hopf technique}

The development of the Wiener-Hopf technique \cite{WH1931} was motivated by what is known as Milne problem \cite{Milne1926}, which consists of a semi-convolution equation,

\begin{equation}
    \int ^{\infty} _ 0 \textbf{H}(t-\tau)\textbf{W}(\tau)d\tau = \textbf{h}(t), \hspace{1cm} 0 < t < \infty
    \label{eq:Milne}
\end{equation}

\noindent where $\mathbf{H}$ and $\mathbf{h}$ are known functions later related to the power spectral density (PSD) associated with sensor and actuator signals, and $\mathbf{W}$ is an unknown function.

Under the Wiener-Hopf approach, the equation above can be extended to negative values of $t$ by writing \cite{zich2014}: 

\begin{equation}
    \int ^{\infty} _ {-\infty} \textbf{H}(t-\tau)\textbf{W}(\tau)\textbf{k}(\tau)d\tau = \textbf{h}(t)\textbf{k}(t) + \textbf{F}(t)\textbf{k}(-t), \hspace{1cm} -\infty<t<\infty
    \label{eq:Milne2}
\end{equation}

\noindent where $\textbf{k}(t)$ is the Heaviside step function and $\textbf{F}(t)$ is an additional unknown function, with the property $\mathbf{F}(t>0)=0$.

Applying Fourier transform to the equation above yields the well known Wiener-Hopf equation \cite{Noble1958}:

\begin{equation}
     \mathbf{\hat{H}}(\omega)\mathbf{\hat{W}}_+(\omega) = \mathbf{\hat{F}}_-(\omega) +\mathbf{\hat{h}}_+(\omega)
     \label{eq:eqWH}
\end{equation}

\noindent with the Fourier transforms of the plus and minus functions being defined respectively as the following half-range Fourier transforms:

\begin{equation}
     \mathbf{\hat{W}_+}(\omega) = \int ^{\infty} _ {0}  \mathbf{\hat{W}}(t)e^{i\omega t}dt, \hspace{1cm}
     \mathbf{\hat{F}_-}(\omega) = \int ^{0} _ {-\infty}  \mathbf{\hat{F}}(t)e^{i\omega t}dt
     \label{eq:pm_def}
\end{equation}

The Wiener-Hopf equations can be solved with multiplicative and additive factorization of the known functions $\mathbf{H}$ and $\mathbf{h}$, defined as:

\begin{equation}
   \mathbf{\hat{H}}(\omega)= \mathbf{\hat{H}_-}(\omega)\mathbf{\hat{H}_+}(\omega),
    \label{eq:Hfac}
\end{equation}

\begin{equation}
   ( \mathbf{\hat{H}}_-^{-1}(\omega)  \mathbf{\hat{h}_+}(\omega))  = ( \mathbf{\hat{H}}_-^{-1}(\omega)  \mathbf{\hat{h}_+}(\omega))_+ + ( \mathbf{\hat{H}}_-^{-1}(\omega)  \mathbf{\hat{h}_+}(\omega))_-.
    \label{eq:Afac}
\end{equation}

The additive factorization of Eq. (\ref{eq:Afac}) will guarantee that the term that carries the $\mathbf{h}$ function will be either a plus or a minus function; thus, the multiplicative factorization of $\mathbf{h}$ is not needed. Application of the factorizations above in the Wiener-Hopf equation and separation of plus and minus terms leads to

\begin{equation}
   \mathcal{L}(\omega) = \mathbf{\hat{H}}_+^{-1}(\omega)\mathbf{\hat{W}}_+(\omega) - ( \mathbf{\hat{H}}_-^{-1}(\omega)  \mathbf{\hat{h}}(\omega))_+ = \mathbf{\hat{H}}_-^{-1}(\omega)\mathbf{\hat{F}}_-(\omega) + ( \mathbf{\hat{H}}_-^{-1}(\omega)  \mathbf{\hat{h}}(\omega))_-
   \label{eq:WHs1}
\end{equation}

Since the part of the equation that carries the minus functions is regular in the lower half of the $\omega$-plane, and the plus term is regular in the upper half, by analytic continuation, $\mathcal{L}(\omega)$ is defined and regular in the whole complex plane. Moreover, it tends to zero as $\omega$ tends to infinity in any direction. Thus, from Liouville's theorem, $\mathcal{L}(\omega) = 0$ \cite{Noble1958}, which yields the following solution:   

\begin{equation}
   \mathbf{\hat{W}}_+(\omega) = \mathbf{\hat{H}}_+^{-1}(\omega)( \mathbf{\hat{H}}_-^{-1}(\omega)  \mathbf{\hat{h}}(\omega))_+ 
   \label{eq:WHs2}
\end{equation}

\begin{equation}
   \mathbf{\hat{F}}_-(\omega) = -\mathbf{\hat{H}}_-(\omega)( \mathbf{\hat{H}}_-^{-1}(\omega)  \mathbf{\hat{h}}(\omega))_+ 
   \label{eq:WHs3}
\end{equation}

For further details about the Wiener-Hopf equation and its solutions, the reader is referred to \citet{Noble1958} and \citet{Martini2022}. The latter reference also details the multiplicative factorization used in this present work, which was based on the method by \citet{daniele2007}.

\newpage
\bibliography{apssamp}

\end{document}